\documentclass[12pt]{article}
\usepackage{a4wide}
\usepackage{epsfig}
\usepackage{amsmath}

\newcommand{\half}{{\textstyle\frac{1}{2}}}

\newlength{\absize}
\catcode`@=11
\def\citer{\@ifnextchar [{\@tempswatrue\@citexr}{\@tempswafalse\@citexr[]}}

\def\@citexr[#1]#2{\if@filesw\immediate
  \write\@auxout{\string\citation{#2}}\fi
  \def\@citea{}\@cite{\@for\@citeb:=#2\do
    {\@citea\def\@citea{--\penalty\@m}\@ifundefined
       {b@\@citeb}{{\bf ?}\@warning
       {Citation `\@citeb' on page \thepage \space undefined}}%
\hbox{\csname b@\@citeb\endcsname}}}{#1}}
\catcode`@=12

\begin{document}
  \thispagestyle{empty}
  \pagestyle{empty}
  \renewcommand{\thefootnote}{\fnsymbol{footnote}}
\newpage\normalsize
    \pagestyle{plain}
    \setlength{\baselineskip}{4ex}\par
    \setcounter{footnote}{0}
    \renewcommand{\thefootnote}{\arabic{footnote}}
\newcommand{\preprint}[1]{%
  \begin{flushright}
    \setlength{\baselineskip}{3ex} #1
  \end{flushright}}
\renewcommand{\title}[1]{%
  \begin{center}
    \LARGE #1
  \end{center}\par}
\renewcommand{\author}[1]{%
  \vspace{2ex}
  {\Large
   \begin{center}
     \setlength{\baselineskip}{3ex} #1 \par
   \end{center}}}
\renewcommand{\thanks}[1]{\footnote{#1}}
\begin{flushright}
{\tt University of Bergen, Department of Physics}    \\[4pt]
{\tt Scientific/Technical Report No.~2000-05}    \\[4pt]
{\tt ISSN 0803-2696} \\[5pt]
{hep-th/0010176} \\[2pt]
October 2000
\end{flushright}
\vskip 0.5cm

\begin{center}
{\large \bf Critical Phenomenon of a Consistent 
$q$-Deformed Squeezed State}
\end{center}
\vspace{1cm}
\begin{center}
P. Osland$^{a,*}$ and Jian-zu Zhang$^{a,b,\#}$
\end{center}
%-----------------------------------
%   Address
%-----------------------------------
\vspace{1cm}
\begin{center}
$^a$ Department of Physics,
University of Bergen, N-5007 Bergen, Norway \\
$^b$ Institute for Theoretical Physics, Box 316,
East China University of Science and Technology,
Shanghai 200237, P. R. China
\end{center}
\vspace{1cm}
%%%%%%%%%%%%%%%%%%%%%%%%%%%%%%%%%%%%%%%%%%%%%%%%%%%%%%%%%%%%%%

\begin{abstract}
Within a self-consistent framework of $q$-deformed Heisenberg algebra 
and its equivalent framework of $q$-deformed boson commutation relations, 
which relate to the under-cutting phenomenon of Heisenberg's minimal 
uncertainty relation, 
special $q$-deformed squeezed states are constructed. 
Besides the similar local maximum squeezing as the one in 
the undeformed case, 
new strong squeezing appears when the amplitude of the related 
coherent state increases to large values. A critical phenomenon appears 
at a large value of the amplitude: 
the variance of one component of the quadrature of the light field 
approaches zero, but the variance of the corresponding conjugate 
quantity remains finite, which is a surprising deviation from 
Heisenberg's uncertainty relation.
The qualitative character exposed by this $q$-squeezed state may provide
some evidence about $q$-deformed effects in current experiments. 

\end{abstract}

\begin{flushleft}
${^*}$ E-mail address: Per.Osland@fi.uib.no \\
${^\#}$ E-mail address: jzzhangw@online.sh.cn
\end{flushleft}

%%%%%%%%%%%%%%%%%%%%%%%%%%%%%%%%%%%%%%%%%%%%%%%%%%%%%%%%%%%%%%%%%%%%%
\section{Introduction}
\setcounter{equation}{0}
%%%%%%%%%%%%%%%%%%%%%%%%%%%%%%%%%%%%%%%%%%%%%%%%%%%%%%%%%%%%%%%%%%%%%        
In discussions of the Heisenberg minimal uncertainty relation special 
attention has been focused on coherent states and squeezed states 
of the light field \cite{Walls}. 
The idea of squeezing has both fundamental and practical 
interests. A special physical proposal for obtaining squeezed state is based 
on the superposition of coherent states along a straight line on 
the plane of $\alpha$, the eigenvalues of the annihilation operators 
\cite{Janszky}, 
which is easy to generalize to the $q$-deformed case. 
Heisenberg's uncertainty relation is a direct result of 
the Heisenberg commutation relation, which is the basis of quantum mechanics. 

According to present tests of quantum electrodynamics, quantum theories 
based on Heisenberg's commutation relation are correct at least 
down to $10^{-18}$~cm. 
The question arises whether there is a possible generalization 
of Heisenberg's commutation relation at shorter distances. 
In searching for such a possibility at short distances considerations 
of the space structure is a useful guide. 
If the space structure at such short distances exhibits 
a non-commutative property, and thus is governed by a quantum group symmetry, 
it has been shown that $q$-deformed quantum mechanics is a 
candidate for a possible pre-quantum theory at short distances. 
In the literature different frameworks of $q$-deformed quantum mechanics 
were established \citer{Janszky,Kempf}. 

At the level of the uncertainty relation there are two kinds of 
modifications of Heisenberg's uncertainty relation 
\cite{Fichtmuller,JZZ98,JZZ99,Codriansky,McDermott,Kempf}, 
implying under- or over-cutting of Heisenberg's minimal uncertainty relation.
Because the framework of $q$-deformed Heisenberg algebra developed 
in Refs.~\cite{Fichtmuller,Hebecker}, which is
associated with the under-cutting phenomenon \cite{JZZ98,JZZ99} shows 
clear physical contents: its relation to the corresponding
$q$-deformed boson commutation relation and the limiting process of
the $q$-deformed harmonic oscillator to the undeformed one are clear, 
in this paper our attention is focused on this framework.
We investigate a special $q$-deformed squeezed state, 
which is a $q$-deformation of the squeezed state proposed in 
Ref.~\cite{Janszky}. 

First we need to clarify consistent representations of $q$-occupation 
number states and $q$-coherent states, which are consistent with the 
$q$-deformed Heisenberg algebra. Then we construct $q$-generalization of the
squeezed states proposed in \cite{Janszky}.

For such a $q$-deformed squeezed state there is a similar 
local maximum of squeezing as in the undeformed case, 
but the state shows an interesting new characteristic:            
As the amplitude of the related coherent state increases beyond the value
corresponding to the local maximum of squeezing,
the variance of one quadrature component of the light field
shows a clear tendency of a second squeezing, 
whereas the corresponding variance for the undeformed case increases to 
the Heisenberg minimal uncertainty. 
Furthermore, the $q$-deformed one exhibits a critical phenomenon: 
when the amplitude of the related coherent state increases to 
a critical point the variance of one component of the quadrature 
of the light field approaches zero; 
meanwhile the variance of the corresponding conjugate operator 
remains finite. 
This is a surprising deviation from Heisenberg's uncertainty relation. 
The qualitative character exposed by this $q$-squeezed state 
is a clear indication of deviation from
the Heisenberg uncertainty relation and
may provide some evidence about $q$-deformed effects to present 
experiments.

In order to demonstrate the critical phenomenon of $q$-deformed squeezed 
states, in the following we first review the necessary background 
of $q$-deformed quantum mechanics.    
%%%%%%%%%%%%%%%%%%%%%%%%%%%%%%%%%%%%%%%%%%%%%%%%%%%%%%%%%%%%%%%%%%%%%
\section{The $q$-deformed Heisenberg Algebra}
\setcounter{equation}{0}
%%%%%%%%%%%%%%%%%%%%%%%%%%%%%%%%%%%%%%%%%%%%%%%%%%%%%%%%%%%%%%%%%%%%%
In terms of 
$q$-deformed phase space variables --- 
the position operator $X$ and the momentum operator $P$, 
the following $q$-deformed Heisenberg algebra is 
developed \cite{Fichtmuller,Hebecker}:
\begin{equation}
\label{Eq:q-algebra}
q^{1/2}XP-q^{-1/2}PX=iU, \qquad
UX=q^{-1}XU, \qquad
UP=qPU,
\end{equation}
where $X$ and $P$ are hermitian and $U$ is unitary:
$X^\dagger=X$, $P^\dagger=P$, $U^\dagger=U^{-1}$.
In (\ref{Eq:q-algebra}) the parameter $q$ is a fixed real number. 
It is important to make distinctions for different realizations of 
the $q$-algebra by different ranges of $q$ values \citer{Zachos,Solomon}. 
Following Refs.~\cite{Fichtmuller,Hebecker} we 
only consider the case $q>1$ in this paper. 
The operator $U$ is called the scaling operator, 
it closely relates to properties of the dynamics and plays an 
essential role in $q$-deformed quantum mechanics. 
The definition of the algebra 
(\ref{Eq:q-algebra}) is based on the definition of the hermitian momentum 
operator $P$. 
However, if $X$ is assumed to be a hermitian operator in a Hilbert space 
the usual quantization rule $P\to -i\partial_X$ does not yield a hermitian 
momentum operator. Ref.~\cite{Fichtmuller} showed that a hermitian 
momentum operator 
$P$ is related to $\partial_X$ and $X$ in a nonlinear way by introducing 
a scaling operator $U$
\begin{equation}
\label{Eq:II-2}
U^{-1}=q^{1/2}[1+(q-1)X\partial_X], \qquad
\bar\partial_X=-q^{-1/2}U\partial_X, \qquad
P=-\frac{i}{2}(\partial_X-\bar\partial_X).
\end{equation}

Such defined hermitian momentum $P$ leads to $q$-deformed effects, 
which are exhibited by the dynamical equation.
For example, the perturbative expansion of the $q$-deformed Schr\"odinger 
equation possesses a complex structure, which amounts to some additional 
momentum-dependent interaction 
\citer{Fichtmuller,LW}, \cite{Hebecker}, \citer{JZZ98,JZZ00}.
The nontrivial properties of  
$U$ imply that the algebra (\ref{Eq:q-algebra}) has a richer structure
than the Heisenberg commutation relation. 
In the limit $q\to 1^+$ the scaling operator $U$ reduces 
to the unit operator, thus the algebra (\ref{Eq:q-algebra}) reduces to 
the Heisenberg commutation relation.  
%%%%%%%%%%%%%%%%%%%%%%%%%%%%%%%%%%%%%%%%%%%%%%%%%%%%%%%%%%%%%%%%%%%%%
\section{The $q$-deformed Boson Commutation Relation.
Wess' Ansatz}
\setcounter{equation}{0}
%%%%%%%%%%%%%%%%%%%%%%%%%%%%%%%%%%%%%%%%%%%%%%%%%%%%%%%%%%%%%%%%%%%%%
Now we consider the general framework of $q$-deformed boson commutation 
relations determined by the properties of the $q$-deformed annihilation, 
creation and number operators: $a$, $a^\dagger$ and $N$. 
The expressions for $a$ and $a^\dagger$ in terms of the $q$-deformed 
variables $X$, $P$, and the scaling operator $U$ (Wess' Ansatz) 
are \cite{LRW}
\begin{equation}
\label{Eq:Wess-Ansatz}
a=\eta U^{-2M}+\beta U^{-M}P, \qquad
a^\dagger=\eta^* U^{2M}+\beta^* PU^{M}
\end{equation}
where $M=1,2,3,\ldots$, and $\eta^*$ and $\beta^*$ are complex 
conjugates of the parameters $\eta$ and $\beta$. 
From Wess' Ansatz (\ref{Eq:Wess-Ansatz}) it follows that $a$ and $a^\dagger$
satisfy the following $q$-deformed boson commutation relation:
\begin{equation}
\label{Eq:a-adagger}
a a^\dagger -q^{-2M} a^\dagger a=(1-q^{-2M})\eta\eta^*=1.
\end{equation}
The right-hand side of (\ref{Eq:a-adagger}) can be normalized to 1 
for the case $q>1$, 
which determines $\eta$ up to a phase: $\eta=e^{i\phi}/(1-q^{-2M})^{1/2}$, 
where $\phi$ is a real number. 

Wess' Ansatz (\ref{Eq:Wess-Ansatz}) is determined by the requirement of  
equivalence of the algebras (\ref{Eq:q-algebra}) and (\ref{Eq:Wess-Ansatz}). 
The operators $a$ and $a^\dagger$ are related 
to the operator $X$ in a complicated way.
(In (\ref{Eq:Wess-Ansatz}) $X$ is nonlinearly included in the operator $U$.)
It is interesting to note that in the limiting case $q\to1^+$
the $q$-deformed annihilation operator $a$ reduces to the undeformed 
one \cite{LRW}. 
The $q$-deformed phase space variables $X$, $P$ and 
the scaling operator $U$ can be realized by 
the variables $\hat x$ and $\hat p$, 
which satisfy $\hat x=\hat x^\dagger$, $\hat p=\hat p^\dagger$
and $[\hat x,\hat p]=i$, as follows \cite{Fichtmuller}:
\begin{equation}
\label{Eq:variables1}
X=\frac{[\hat z+\half]}{\hat z+\half}\,\hat x, \qquad
P=\hat p, \qquad
U=q^{\hat z},
\end{equation}
where $\hat z=-i(\hat x\hat p+\hat p\hat x)/2$, and
$[A]$ is called the $q$-deformation of $A$, defined by 
$[A]=(q^A-q^{-A})/(q-q^{-1})$.
From (\ref{Eq:variables1}) it follows that $X$, $P$ and $U$ satisfy 
(\ref{Eq:q-algebra}). 
Furthermore, $\hat x$ and $\hat p$ are realized by conventional variables 
$x$ and $p$, 
\begin{equation}
\label{Eq:realization}
\hat x=x, \qquad \hat p=p+\gamma(1-q^{-2M})^{-1/2}, \qquad
p=-i\partial_x,
\end{equation}
where $\gamma$ is a real parameter. 

In order to study the behavior of $a$ in the limit 
$q\to1^+$, let $q=e^f$, where $0<f\ll1$. In the limit $q\to1^+$
($f\to0^+$) there are singular factors in $\eta$ and $\hat p$. 
The condition of cancellation of these two singular terms is 
$\beta\gamma=-e^{i\phi}$. If we take $e^{i\phi}=\mp i$, 
$\gamma=\pm(2\omega)^{1/2}$, $\beta=i(2\omega)^{-1/2}$, 
where the constant $\omega$ is the frequency of the ordinary oscillator, 
then in the limit $f\to0^+$ the $q$-deformed annihilation operator $a$  
reduces to the ordinary annihilation 
operator: $a\to a_0=(\omega/2)^{1/2}x+i(2\omega)^{-1/2}p$. 
The $q$-deformed harmonic oscillators were first studied by 
Macfarlane \cite{Macfarlane} and Biedenharn \cite{Biedenharn}.

The definition of the $q$-deformed number operator $N$ is the same as 
for the undeformed one: $[N,a]=-a$, $[N,a^\dagger]=a^\dagger$.
But the relations between $a^\dagger a$ ($aa^\dagger$)
and $N$ are involved. From (\ref{Eq:a-adagger}) it follows that
(in the following we only consider the case $M=1$)
\begin{equation}
\label{Eq:II-4}
a^\dagger a=q^{-(N-1)}[N], \qquad
a a^\dagger=q^{-N}[N+1],
\end{equation}
The eigenstate $|n\rangle_q$ of $N$ satisfies  
$N|n\rangle_q=n|n\rangle_q$, 
with ${}_q\langle n|m\rangle_q=\delta_{nm}$ and 
\begin{equation}
\label{Eq:II-5}
a|n\rangle_q=q^{-(n-1)/2}[n]^{1/2}|n-1\rangle_q, \qquad
a^\dagger|n\rangle_q=q^{-n/2}[n+1]^{1/2}|n+1\rangle_q.
\end{equation}
The $q$-deformed vacuum $|0\rangle_q$ satisfies $a|0\rangle_q=0$.

We emphasize that the $q$-exponential factors in (\ref{Eq:II-4}) 
and (\ref{Eq:II-5}) are delicate points in applications of 
$q$-deformed oscillators. 
From (\ref{Eq:II-5}) it follows that the $q$-occupation-number state is 
\begin{equation}
\label{Eq:II-6}
|n\rangle_q=\frac{q^{n(n-1)/4}}{\sqrt{[n]!}}
(a^\dagger)^n|0\rangle_q.
\end{equation}
where $[n]!=[1][2]\cdots[n]$, and $[0]\equiv1$. 

As an application of the $q$-occupation number state (\ref{Eq:II-6}) 
we consider the eigenvalues of the $q$-deformed Hamiltonian. 
In the literature there are two $q$-deformed Hamiltonians 
\cite{Fichtmuller,LRW,JZZ99,Nelson}: 
$H_\omega=\omega(a^\dagger a+\half)$ and 
$H_{Q,K}=\half K^2+\half \omega^2Q^2$,  
where $Q$ and $K$ are a pair of quadratures
of $a$ and $a^\dagger$:
\begin{equation}
\label{Eq:II-7}
Q=\sqrt{\frac{1}{2\omega}}(a+a^\dagger), \qquad
K=i\sqrt{\frac{\omega}{2}}(a^\dagger-a).
\end{equation}
Refs.~\cite{Fichtmuller,LRW,JZZ99,Nelson}
note the difference between $H_\omega$ and 
$H_{Q,K}$ that $H_\omega$ possesses conventional physical properties, 
but $H_{Q,K}$ probably does not permit a consistent physical interpretation. 
Thus we only consider the eigenvalues of $H_\omega$. 
From (\ref{Eq:II-4}) it follows that $|n\rangle_q$ is an eigenstate
of $H_\omega$ with eigenvalue
\begin{equation}
\label{Eq:II-8}
E_{\omega,n}=\omega(q^{-(n-1)}[n]+\half)
=\omega[(1-q^{-2n})/(1-q^{-2})+\half].
\end{equation}
In the limit $q\to1^+$, the spectrum $E_{\omega,n}$ in (\ref{Eq:II-8}) 
is reduced 
to the undeformed one $E_{\omega,n}=\omega(n+\half)$. 
We also note that in the limit $n\to\infty$
the eigenvalues $E_{\omega,n}$ are bounded from above, 
$E_{\omega,\infty}=\omega[(1-q^{-2})^{-1}+\half]$. 
The spectrum $E_{\omega,n}$ in (\ref{Eq:II-8}) is identified with the bounded 
spectrum of Ref.~\cite{LRW}.
%%%%%%%%%%%%%%%%%%%%%%%%%%%%%%%%%%%%%%%%%%%%%%%%%%%%%%%%%%%%%%%%%%%%%
\section{Undercutting of Heisenberg's Minimal 
Uncertainty Relation}
\setcounter{equation}{0}
%%%%%%%%%%%%%%%%%%%%%%%%%%%%%%%%%%%%%%%%%%%%%%%%%%%%%%%%%%%%%%%%%%%%%

The $q$-deformed uncertainty relation derived from the
$q$-deformed boson commutation relations (\ref{Eq:a-adagger}) 
differs qualitatively from that of Heisenberg \cite{JZZ98,JZZ99}.
In order to expose such characteristic 
quantum behavior governed by (\ref{Eq:a-adagger}) we proceed as follows. 
From (\ref{Eq:a-adagger}) (with $M=1$) and (\ref{Eq:II-7}) 
it follows that the commutation relation between $Q$ and $K$ is:
\begin{equation}
\label{Eq:IV-1}
[Q,K]=iC, \qquad C=1-(1-q^{-2})a^\dagger a.
\end{equation}
It was shown that in any state the expectation value of the operator $C$ 
satisfies \cite{JZZ98,JZZ99}
\begin{equation}
\label{Eq:IV-2}
0\le \langle C\rangle \le1.
\end{equation}
In (\ref{Eq:IV-2}) the situation $\langle C\rangle=1$ only occurs 
for the case $q=1$, which means that Heisenberg's minimal uncertainty 
relation $\Delta Q\Delta K=\half$ is undercut for $q>1$, where  
$\Delta A\equiv\langle(A-\bar A)^2\rangle^{1/2}$, 
and $\bar A\equiv\langle A\rangle$. 
It is quite interesting to note that the possibility $\langle C\rangle=0$ 
can occur. In this case the $q$-deformed 
uncertainty relation differs essentially from that of Heisenberg. It 
means that there are special states, in which $\Delta Q$ and $\Delta K$ 
can be simultaneously zero. 
Within the $q$-deformed coherent state, such an example has been
found \cite{JZZ99,McDermott}.

%%%%%%%%%%%%%%%%%%%%%%%%%%%%%%%%%%%%%%%%%%%%%%%%%%%%%%%%%%%%%%%%%%%%%
\section{$q$-deformed Coherent State \cite{Nelson}}
\setcounter{equation}{0}
%%%%%%%%%%%%%%%%%%%%%%%%%%%%%%%%%%%%%%%%%%%%%%%%%%%%%%%%%%%%%%%%%%%%%

Using (\ref{Eq:a-adagger}), (\ref{Eq:II-4})--(\ref{Eq:II-6}) 
we construct $q$-coherent states, which are defined as eigenstates of 
the $q$-annihilation operator $a$, 
$a|\alpha\rangle_q=\alpha|\alpha\rangle_q$ with a complex eigenvalue 
$\alpha$. 
From the $q$-deformed boson commutation relations (\ref{Eq:a-adagger}) 
it follows that the $q$-coherent state $|\alpha\rangle_q$ is represented as
\begin{equation}
\label{Eq:V-1}
|\alpha\rangle_q=N_\alpha
\left[ |0\rangle_q
+\sum_{n=1}^\infty \frac{q^{n(n-1)/4}\alpha^n}
                        {\sqrt{[n]!}}
|n\rangle_q \right]
=N_\alpha e_q(\alpha a^\dagger)|0\rangle_q,
\end{equation}
where the function $e_q(x)$ is 
the $q$-deformed exponential function, which is defined as
\begin{equation}
\label{Eq:V-2}
e_q(x)=1+\sum_{n=1}^\infty
\frac{q^{n(n-1)/2}}{[n]!}\, x^n.
\end{equation}
When $q\to1^+$,  
$e_q(x)\to\exp(x)$.  
We emphasize again that the $q$-exponential factors in 
(\ref{Eq:II-4})--(\ref{Eq:II-6}), (\ref{Eq:V-1}) and (\ref{Eq:V-2}) 
guarantee the consistency of the theory. 
Note that there is an overlap between $|\alpha\rangle_q$ 
and $|-\alpha\rangle_q$:
$_q\langle\alpha|-\alpha\rangle_q=N_{-\alpha}^*N_\alpha\,e_q(-\alpha^2)$.
For simplicity we may choose the phase factor so that $\alpha$ is real, 
thus in (\ref{Eq:V-1}) the normalization constant 
is $N_\alpha=[e_q(\alpha^2)]^{-1/2}$.
%%%%%%%%%%%%%%%%%%%%%%%%%%%%%%%%%%%%%%%%%%%%%%%%%%%%%%%%%%%%%%%%%%%%%
\section{Critical Phenomenon of $q$-deformed Squeezed State}
\setcounter{equation}{0}
%%%%%%%%%%%%%%%%%%%%%%%%%%%%%%%%%%%%%%%%%%%%%%%%%%%%%%%%%%%%%%%%%%%%%
We now consider a special $q$-squeezed state, 
which is a $q$-deformation of the 
squeezed state considered in Ref. \cite{Janszky}. An effective squeezing 
can be achieved by superposition of coherent states along a straight line 
on the $\alpha$ plane. 
For a single mode of frequency $\omega$ the electric field operator 
$E(t)$ is represented as 
$E(t)=E_0[a\exp(-i\omega t)+a^\dagger\exp(i\omega t)]$,
where $a$ and $a^\dagger$ are the 
annihilation and creation operators of photon field.
This squeezed state is defined as 
\begin{equation}
\label{Eq:VI-1}
|\alpha,\pm\rangle_q
=c_\pm\left(|\alpha\rangle_q\pm|-\alpha\rangle_q\right),
\end{equation}
which satisfies
$a|\alpha,\pm\rangle_q=\alpha c_\pm c_\mp^{-1}|\alpha,\mp\rangle_q$,
$a^2|\alpha,\pm\rangle_q=\alpha^2|\alpha,\pm\rangle_q$, and  
${}_q\langle\alpha,\pm|\alpha,\mp\rangle_q=0$. 
The normalization constants are
$c_\pm^2=e_q(\alpha^2)/2[e_q(\alpha^2)\pm e_q(-\alpha^2)]$.

In the state $|\alpha,+\rangle_q$ squeezing appears. 
Let $a_1$ and $a_2$ be the dimensionless hermitian quadratures of the 
annihilation operator: $a=a_1+ia_2$.
The variances of $a_1$ and
$a_2$ in this state are:
\begin{eqnarray}
\label{Eq:VI-2}
(\Delta a_1)^2 &=& \frac{1}{4}+\frac{\alpha^2}{4}
\left[2+(1+q^{-2})\frac{e_q(\alpha^2)-e_q(-\alpha^2)}
                      {e_q(\alpha^2)+e_q(-\alpha^2)} \right], \\
\label{Eq:VI-3}
(\Delta a_2)^2 &=& \frac{1}{4}-\frac{\alpha^2}{4}
\left[2-(1+q^{-2})\frac{e_q(\alpha^2)-e_q(-\alpha^2)}
                      {e_q(\alpha^2)+e_q(-\alpha^2)} \right].
\end{eqnarray}
The state $|\alpha,+\rangle_q$ is squeezed, i.e.,
the variance $(\Delta a_2)^2$ is less than Heisenberg's 
minimal uncertainty 0.25. 
When $q\to1^+$, $(\Delta a_2)^2$ reduces to the undeformed value 
\cite{Janszky} 
\begin{equation}
(\Delta a_2)_{\rm un}^2 = \frac{1}{4}-\frac{\alpha^2}{1+\exp(2\alpha^2)}.
\end{equation}
For $(\Delta a_2)_{\rm un}^2$, there is only one maximum squeezing,
which appears at $\alpha_{\rm un}^2=0.64$, 
where $(\Delta a_2)_{\rm un}^2=0.111$.
Beyond $\alpha_{\rm un}^2$, as $\alpha^2$ increases, 
$(\Delta a_2)_{\rm un}^2$ monotonically increases to Heisenberg's 
minimal uncertainty 0.25.

For the $q$-deformed case, like the undeformed case,
there is a similar local maximum squeezing.
We consider a few numerical examples:
$f=0.0001$, 0.001 and 0.01;
the corresponding values for $(\Delta a_2)^2$ are shown in Fig.~1.
The local maximum squeezing $(\Delta a_2)^2\simeq 0.11$ appears at 
$\alpha_{\rm q-def}^2\simeq 0.64$. 

But (\ref{Eq:VI-3}) shows an essential new characteristic. 
In order to expose such new characteristic we proceed as follows.

%%%%%%%%%%%%%%%%%%%%%%%%%%%%%%%%%%%%%%%%%%%%%%%%%%%%%%%%%%%%%%%%%%%%%%%%
\begin{figure}[htb]
\refstepcounter{figure}
\label{Fig:onedim}
\addtocounter{figure}{-1}
\begin{center}
\setlength{\unitlength}{1cm}
\begin{picture}(8.0,8.0)
\put(0,-0.3)
{\mbox{\epsfysize=8.5cm\epsffile{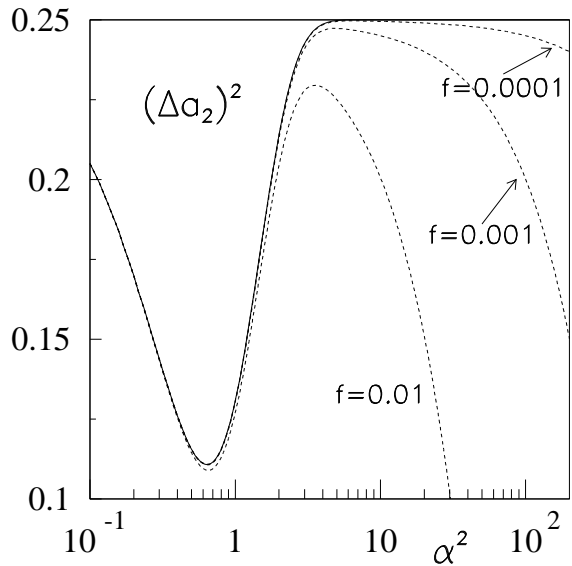}}}
\end{picture}
\caption{Variance $(\Delta a_2)^2$ {\it vs.}\ $\alpha^2$, for three
values of $f$, where $q=e^f$ (dashed), and for the undeformed
case (solid).}
\end{center}
\end{figure}
%%%%%%%%%%%%%%%%%%%%%%%%%%%%%%%%%%%%%%%%%%%%%%%%%%%%%%%%%%%%%%%%%%%%%%%%

The behavior of $\Delta a_2$ depends on the value of $q$.
In the limit $q\to 1^+$ ($f\to 0^+$) this behaviour can be studied
analytically. Since the series expansion for the $q$-exponential, 
Eq.~(\ref{Eq:V-2}), is absolutely convergent, we may differentiate
term by term, and find
\begin{equation}
e_q(x)
\simeq\left[1+\half(q-1)x^2\right]e^x, \qquad \text{as }q\to1^+.
\end{equation}
Substituting this into Eq.~(\ref{Eq:VI-3}), we find
\begin{equation}
\label{Eq:deltasq2-asympt}
(\Delta a_2)^2 \simeq
\frac{1}{4}-\frac{\alpha^2}{4}
\left[2-(1+q^{-2})\tanh(\alpha^2)\right].
\end{equation}
For $q\to1^+$, the critical value can be determined from 
Eq.~(\ref{Eq:deltasq2-asympt}). 

For large $\alpha^2$ it follows that
$(\Delta a_2)^2 = \frac{1}{4}(1-2\alpha^2 f)$.
Thus, the large $\alpha^2$-behavior of $(\Delta a_2)^2$ is quite different
from that of the undeformed $(\Delta a_2)_{\rm un}^2$:
As $\alpha^2$ increases to large values, $(\Delta a_2)^2$ shows 
further strong squeezing; 
and at the point $\alpha_c^2=1/(2f)$ a critical squeezing appears: 
we obtain a zero variance $(\Delta a_2)^2=0$,
meanwhile $(\Delta a_1)^2=1/(2f)$ remains finite.
In Fig.~2 we show the $q$- and $\alpha^2$-dependence of $(\Delta a_2)^2$
from Eq.~(\ref{Eq:VI-3}).
%%%%%%%%%%%%%%%%%%%%%%%%%%%%%%%%%%%%%%%%%%%%%%%%%%%%%%%%%%%%%%%%%%%%%%%%
\begin{figure}[htb]
\refstepcounter{figure}
\label{Fig:twodim}
\addtocounter{figure}{-1}
\begin{center}
\setlength{\unitlength}{1cm}
\begin{picture}(8.0,8.0)
\put(0,-0.3)
{\mbox{\epsfysize=8.5cm\epsffile{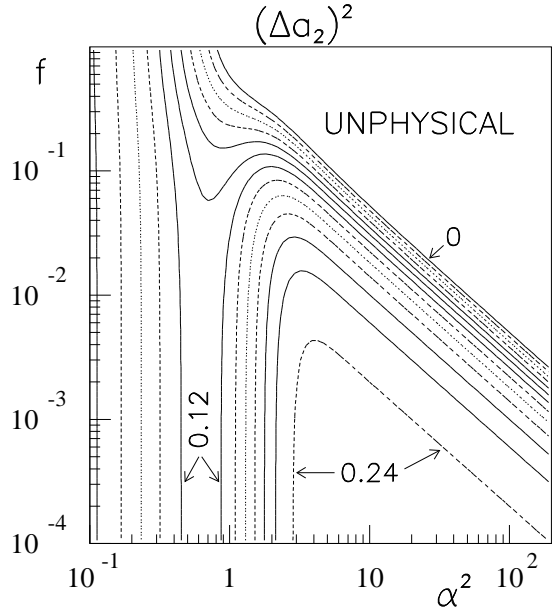}}}
\end{picture}
\caption{Variance $(\Delta a_2)^2$ {\it vs.}\ $\alpha^2$ and $f$.
Contours of $(\Delta a_2)^2$ are drawn at values $0, 0.02, \ldots, 0.24$.
The region to the upper right is unphysical.}
\end{center}
\end{figure}
%%%%%%%%%%%%%%%%%%%%%%%%%%%%%%%%%%%%%%%%%%%%%%%%%%%%%%%%%%%%%%%%%%%%%%%%
At low $f$, there is a local minimum around $\alpha^2\sim0.64$,
representing squeezing, followed by a plateau where $(\Delta a_2)^2$
approaches the Heisenberg value of 0.25.
Eventually, though, for yet larger values of $\alpha^2$,
a second squeezing sets in, leading to a critical point where
$(\Delta a_2)^2$ vanishes.

It should be emphasized that $q$-deformed quantum mechanics 
is essentially different from ordinary quantum mechanics. 
It shows qualitative deviations from Heisenberg's minimal 
uncertainty relation.
For example, there are some special states, which permit simultaneously 
zero minimal uncertainties in a pair of conjugate operators 
\cite{JZZ98,JZZ99}. 
The $q$-deformed coherent state \cite{JZZ99,McDermott} is such an example. 
Here, we have discussed another case, the special $q$-squeezed state,
which shows another qualitative deviation from Heisenberg's 
uncertainty relation.
In a pair of conjugate operators, when one variance approaches zero, 
the other variance still remains finite, 
which is a surprising deviation from Heisenberg's uncertainty relation:
for quantum mechanics if one variance approaches zero the variance 
of the conjugate operator should approach infinity. 

It should be pointed out that if $q$-deformed quantum mechanics is 
a relevant physical theory, then its effects mainly manifest themselves
at very short distances much smaller than $10^{-18}$~cm; 
its correction to ordinary quantum mechanics must be very small in the 
energy range of present experiments, which means that the parameter $q$
must be very close to one. The critical phenomenon exposed in the 
$q$-squeezed state (\ref{Eq:VI-1}) gives an example, 
which is a clear indication of $q$-deformed effects and may provide 
some evidence about such effects to present experiments. 
The above analysis suggests that in experiments one should consider 
a light field with large $\alpha^2$ and examine the large-$\alpha^2$ 
behavior of the variance of the $a_2$ component.
\vspace{10mm}

This work has been supported by the Research Council of Norway.
JZZ would like to thank the Department of Physics, University of Bergen 
for hospitality. 
His work has also been supported by the National Natural Science 
Foundation of China, and by the Shanghai Education Development 
Foundation.

%%%%%%%%%%%%%%%%%%%%%%%%%%%%%%%%%%%%%%%%%%%%%%%%%%%%%%%%%%%%%%%%%%%%%%%%
%\clearpage

\end{document}